\documentclass[10pt, aps, prl, reprint, superscriptaddress, longbibliography]{revtex4-1} 
\usepackage[utf8]{inputenc}
\usepackage{dcolumn} 
\usepackage{bm} 
\usepackage{amsmath} 
\usepackage{amssymb}
\usepackage{xr}
\usepackage{graphicx}
\usepackage{gensymb} 
\usepackage{natbib}
\usepackage{url}
\usepackage{textcase}
\usepackage{amsfonts}
\usepackage{lineno}
\usepackage{color}

\begin{document}


\title{Nanoscale subsurface dynamics of solids upon high-intensity femtosecond laser irradiation observed by grazing-incidence x-ray scattering}

\author{Lisa Randolph}
\affiliation{Department Physik, Universit\"{a}t Siegen, 57072, Siegen, Germany}
\author{Mohammadreza Banjafar}
\affiliation{European XFEL, 22869, Schenefeld, Germany}
\affiliation{Technical University Dresden, 01069 Dresden, Germany}

\author{Thomas R. Preston}
\affiliation{European XFEL, 22869, Schenefeld, Germany}

\author{Toshinori Yabuuchi}
\affiliation{Japan Synchrotron Radiation Research Institute (JASRI), Sayo, Hyogo, 679-5198, Japan}
\affiliation{RIKEN SPring-8 Center, Sayo, Hyogo 679-5148, Japan}

\author{Mikako Makita}
\affiliation{European XFEL, 22869, Schenefeld, Germany}

\author{Nicholas P. Dover}
\affiliation{QST-Kansai, KPSI, Kyoto, 619-0215, Japan }

\author{Christian R\"{o}del}
\affiliation{Technical University Darmstadt, 64289 Darmstadt, Germany}

\author{Sebastian G\"{o}de}
\affiliation{European XFEL, 22869, Schenefeld, Germany}

\author{Yuichi Inubushi}
\affiliation{Japan Synchrotron Radiation Research Institute (JASRI), Sayo, Hyogo, 679-5198, Japan}
\affiliation{RIKEN SPring-8 Center, Sayo, Hyogo 679-5148, Japan}

\author{Gerhard Jakob}
\affiliation{Johannes Gutenberg-Universit\"{a}t Mainz,55128 Mainz, Germany}

\author{Johannes Kaa}
\affiliation{European XFEL, 22869, Schenefeld, Germany}

\author{Akira Kon}
\affiliation{QST-Kansai, KPSI, Kyoto, 619-0215, Japan }

\author{James K. Koga}
\affiliation{QST-Kansai, KPSI, Kyoto, 619-0215, Japan }

\author{Dmitriy Ksenzov}
\affiliation{Department Physik, Universit\"{a}t Siegen,  57072, Siegen, Germany}

\author{Takeshi Matsuoka}
\affiliation{Open and Transdisciplinary Research Institute, Osaka University, Suita, Osaka 565-0087, Japan}
\affiliation{Graduate School of Engineering, Osaka University, Suita, Osaka 565-0087, Japan}

\author{Mamiko Nishiuchi}
\affiliation{QST-Kansai, KPSI, Kyoto, 619-0215, Japan }

\author{Michael Paulus}
\affiliation{Fakult\"{a}t Physik/DELTA, TU Dortmund, 44221 Dortmund, Germany}

\author{Frederic Schon}
\affiliation{Department Physik, Universit\"{a}t Siegen, 57072, Siegen, Germany}

\author{Keiichi Sueda}
\affiliation{RIKEN SPring-8 Center, Sayo, Hyogo 679-5148, Japan}

\author{Yasuhiko Sentoku}
\affiliation{Institute of Laser Engineering, Osaka University, Suita, Osaka 565-0871, Japan}

\author{Tadashi Togashi}
\affiliation{Japan Synchrotron Radiation Research Institute (JASRI), Sayo, Hyogo, 679-5198, Japan}
\affiliation{RIKEN SPring-8 Center, Sayo, Hyogo 679-5148, Japan}

\author{Mehran Vafaee-Khanjani}
\affiliation{Johannes Gutenberg-Universit\"{a}t Mainz,55128 Mainz, Germany}

\author{Michael Bussmann}
\affiliation{Helmholtz-Zentrum Dresden-Rossendorf, 01328, Dresden, Germany}
\affiliation{Center for Advanced Systems Understanding (CASUS), G\"{o}rlitz, Germany}

\author{Thomas E. Cowan}
\affiliation{Technical University Dresden, 01069 Dresden, Germany}
\affiliation{Helmholtz-Zentrum Dresden-Rossendorf, 01328, Dresden, Germany}

\author{Mathias Kl\"{a}ui}
\affiliation{Johannes Gutenberg-Universit\"{a}t Mainz,55128 Mainz, Germany}

\author{Carsten Fortmann-Grote}
\affiliation{European XFEL, 22869, Schenefeld, Germany}

\author{Lingen Huang}
\affiliation{Helmholtz-Zentrum Dresden-Rossendorf, 01328, Dresden, Germany}

\author{Adrian P. Mancuso}
\affiliation{European XFEL, 22869, Schenefeld, Germany}
\affiliation{Department of Chemistry and Physics, La Trobe Institute for Molecular Science, La Trobe University, Melbourne, Victoria 3086, Australia}

\author{Thomas Kluge}
\affiliation{Helmholtz-Zentrum Dresden-Rossendorf, 01328, Dresden, Germany}

\author{Christian Gutt}
\email{christian.gutt@uni-siegen.de}
\affiliation{Department Physik, Universit\"{a}t Siegen, 57072, Siegen, Germany}

\author{Motoaki Nakatsutsumi}
\email{motoaki.nakatsutsumi@xfel.eu}
\affiliation{European XFEL, 22869, Schenefeld, Germany}
\affiliation{Open and Transdisciplinary Research Institute, Osaka University, Suita, Osaka 565-0087, Japan}

\date{\today}

\begin{abstract}
Observing ultrafast laser-induced structural changes in nanoscale systems is essential for understanding the dynamics of intense light-matter interactions.  
For laser intensities on the order of $10^{14} \, \rm W/cm^2$, highly-collisional plasmas are generated at and below the surface. Subsequent transport processes such as heat conduction, electron-ion thermalization, surface ablation and resolidification occur at picosecond and nanosecond time scales. Imaging methods, e.g. using x-ray free-electron lasers (XFEL), were hitherto unable to measure the depth-resolved subsurface dynamics of laser-solid interactions with appropriate temporal and spatial resolution. Here we demonstrate picosecond grazing-incidence small-angle x-ray scattering (GISAXS) from laser-produced plasmas using XFEL pulses. Using multi-layer (ML) samples, both the surface ablation and subsurface density dynamics are measured with nanometer depth resolution. 
Our experimental data challenges the state-of-the-art modeling of matter under extreme conditions and opens new perspectives for laser material processing and high-energy density science.
\end{abstract}

\maketitle 

\noindent
With the advent of the Chirped Pulse Amplification technique~\cite{Strickland1985}, intensities ranging from $10^{13}\,\rm W/cm^2$ to $10^{23} \, \rm W/cm^2$ \cite{Yoon21} can be achieved by compressing laser pulses to a few femtoseconds and focusing them to small spot sizes. The intensity regime around $10^{14}\,\rm W/cm^2$ is particularly interesting for studies of atomic and molecular physics~\cite{Brabec2000} but also for laser material processing~\cite{Phillips2015, Bonse20}. When a laser pulse is focused on a solid surface at these intensities, the atoms at the surface are ionized and a dense plasma is created. The laser field interacts with electrons within a depth of a few tens of nanometers -- the so-called skin depth -- due to the shielding of the laser field by high-frequency plasma oscillations. Several physical processes occur after the laser pulse on a picosecond to nanosecond time scale~\cite{Rethfeld2017} leading to thermalization, diffusion, compression and ablation which finally results in a new surface structure after resolidification~\cite{Phillips2015}. A good understanding and control of these processes enables high-precision material processing and functional surfaces production~\cite{Bonse20}. These processes generally take place in the regime of strongly-coupled plasmas, where atomic-scale collisions and quantum effects play an important role. From a simulation point of view, implementing both internal atomic dynamics and the long-range light propagation poses a significant challenge \cite{Varin2014, *Bart2017}. Therefore, various assumptions and approximations are involved in state-of-the-art simulation codes that need to be tested with experimental data. 
At laser intensities above $10^{16} \, \rm W/cm^2$, on the other hand, high-density plasmas act as active optical devices allowing manipulation of the temporal \cite{Doumy04} and spatial~\cite{Nakatsutsumi10, Vincenti19} properties of the laser radiation, as well as generating coherent attosecond pulses in the extreme ultraviolet~\cite{Wheeler12}. Furthermore, the interaction of such high intensity lasers with solids has attracted considerable interest in the prospect of creating bright and compact particles~\cite{Daido12, *Macchi13} and gamma-ray sources~\cite{Stark16}. In these experiments, the surface density profile is often perturbed before the arrival of the laser pulse due to imperfect temporal contrast, or during the pulse via its extreme light pressure~\cite{Wilks1992}. These interactions modify the transient optical properties of surface plasmas and influence the laser-plasma coupling mechanisms~\cite{Chopineau19}.  
To summarize, having the capability of \textit{in situ} visualization of the surface \textit{and} subsurface density profile with skin-depth resolution may revolutionize our understanding of laser matter interactions and foster numerous applications. 

So far, surface plasma dynamics have been widely investigated using femtosecond optical laser pulses~\cite{Sokolowski-Tinten1998}, whereby various surface-sensitive methods based on optical interferometry~\cite{Geindre94, *Bocoum2015} and spectroscopy~\cite{Malvache13} have been applied. For resolving physical processes \textit{below} the surface and inside the bulk, x-ray pulses are more suitable as they can penetrate dense plasmas. Experiments using ultrafast x-ray diffraction have revealed, for example, non-thermal melting~\cite{Siders1999}, coherent lattice vibrations~\cite{Sokolowski-Tinten2003}, and ultrafast phase transitions~\cite{Cavalleri2001}. Ultrafast deformation within a thin metal film has been studied using laser-driven soft x-ray sources~\cite{Tobey2007}. Recently, x-ray pulses from XFELs have been applied to investigate laser produced plasmas with femtosecond temporal resolution~\cite{Fletcher2015}. Small-angle scattering of femtosecond x-rays has revealed density gradients of expanding solid-density plasmas with nanometer spatial and femtosecond temporal resolution~\cite{Gorkhover16, *Kluge18}. However, the above methods generally lack depth resolution of the subsurface dynamics. Therefore, absorption and phase contrast x-ray imaging~\cite{Schropp15, *Rigon21} has been developed, which can visualize the dynamics of the bulk density. Despite the large progress in the past few years, the spatial resolution of these methods is limited to about $100\,\rm nm$ and is thus larger than the skin depth. Therefore, x-ray imaging methods are so far unable to resolve small density perturbations on the nanometer scales inherent to laser-driven plasmas.

Here we report the first demonstration of ultrafast grazing-incidence small-angle x-ray scattering (GISAXS) using an XFEL. Measuring the non-specular diffuse x-ray scattering patterns allows us to access different depths inside strongly-coupled dense plasmas with nanometer resolution. This method provides excellent sensitivity for measuring the surface and subsurface density distribution, which facilitates a comparison of complex physics models of laser-driven dense plasmas.

\begin{figure}[hbt!]
    \centering
    \includegraphics[width=\linewidth]{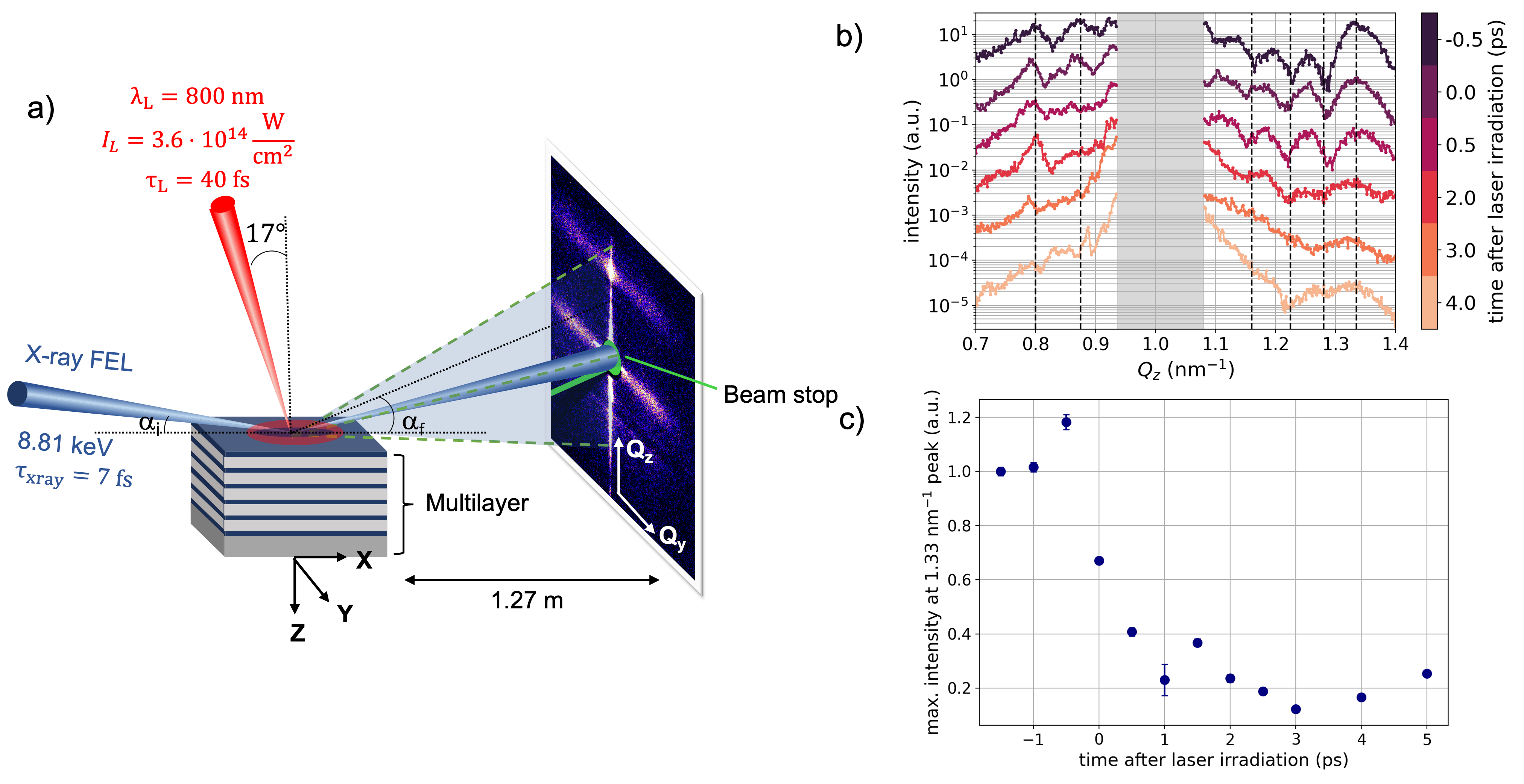}
    \caption{\textbf{Schematics of the experimental setup to investigate the subsurface plasma dynamics of a solid with grazing incidence small-angle x-ray scattering (GISAXS) by a single femtosecond x-ray FEL pulse.} \textbf{a)} The multilayer (ML) samples consisted of 5 layers of tantalum (Ta) and copper nitride ($\mathrm{Cu_3 N}$), of 4.3 / 11.5~nm thickness each. The samples were irradiated by an optical laser with 800~nm central wavelength, $3.6 \pm 0.2 \times 10^{14}\,\mathrm{W/cm}^2$ intensity, 40~fs duration in full-width at half-maximum (FHWM)) under an incident angle of 17$\degree$ from the surface normal in p-polarization. The x-ray pulses with 8.81~keV photon energy, 7~fs FWHM duration, and 4~µm spot size (FWHM) irradiate the sample at the grazing incidence angle of $\alpha_i=0.64\degree$, \textit{i.e.} slightly above the critical angle of external total reflection of the layer materials. The laser beam is defocussed to obtain a $\sim$500~$\mathrm{\mu}$m spot diameter to cover the x-ray footprint of $\sim$360~$\mathrm{\mu}$m at the surface. Scattered x-ray photons are recorded by an MPCCD area detector placed around the specular direction. The strong specular peak at $Q_z = 1.0\,\mathrm{nm}^{-1}$ is blocked.
    \textbf{b)} In-plane signal along $Q_z$ for different time delays between $-0.5$ to $4 \,\mathrm{ps}$ after the laser intensity peak. The position of the beam stop is indicated by the grey shaded area. \textbf{c)} Maximum intensity of the intense peak at $Q_z=1.33 \,\mathrm{nm}^{-1}$ as a function of the time delay.}
    \label{fig:fig1}
\end{figure}

\begin{figure*}[htb!]
    \centering
    \includegraphics[width=\linewidth]{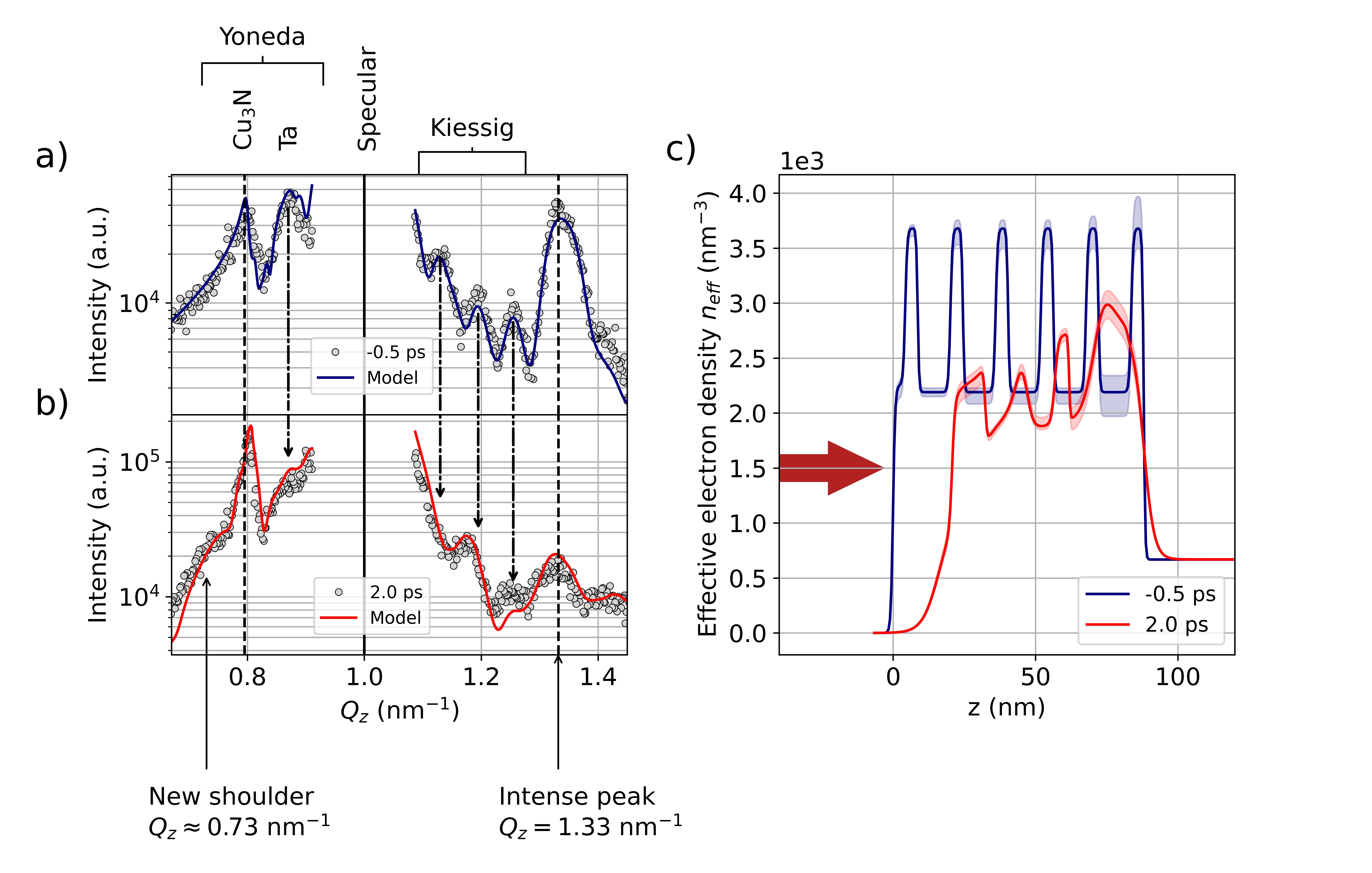}
    \caption{\textbf{GISAXS signals and corresponding retrieved real-space effective electron density profiles.} Lineout of the in-plane scattering at \textbf{a)} 0.5~ps before and \textbf{b)} 2~ps after the laser intensity peak. Solid lines represent refinements using the program BornAgain. \textbf{c)} Retrieved real-space effective electron density profile as a function of the depth ($z$) including its confidence interval (light blue and red, for details see \textit{Supplementary Information}). The red arrow indicates the direction of laser irradiation. 
    }
    \label{fig:fig2}
\end{figure*}
\section{Results}
\subsection{GISAXS analysis}
The experiment was performed at the EH6 station at the SACLA XFEL facility in Japan~\cite{Yabuuchi19}. A metallic multilayer (ML) sample consisted of 5 layers of tantalum (Ta) and copper nitride ($\mathrm{Cu_3 N}$), of 4.3 / 11.5~nm thickness each, was irradiated at 17$\degree$ from the normal to its surface by an optical laser with a central wavelength of 800~nm, intensity $3.6 \pm 0.2 \times 10^{14}\,\mathrm{W/cm}^2$ and pulse duration of 40~fs (Fig.~1~a). Using a variable delay, the sample was irradiated with x-rays of photon energy 8.81~keV at a grazing incidence angle of $\alpha_i=0.64\degree$, i.e., slightly above the total external reflection angles. The scattered x-ray photons around the specular reflection were recorded by an MPCCD area detector~\cite{Kameshima14}. The strong specular peak, where the incident angle equals the exit angle  ($Q_{\text{specular}} = 1.0 \, \mathrm{nm}^{-1}$), was blocked by a beam stop as it is many orders of magnitude more intense than that of the scattered signal.
Due to a large illumination area with the grazing incidence ($\sim 360 \times 4 \, \mathrm{\mu m^2}$), the detected x-ray signal represented a temporal integration of about 1.2~ps. A more detailed description of the setup is provided in the \textit{Methods}. 
Fig. \ref{fig:fig1}~b) shows the time-dependence of the x-ray intensity within the scattering plane -- so called in-plane data at $Q_y=0$. 
Because of the GISAXS geometry, the in-plane scattering signal contains momentum transfers of both normal ($Q_z$) and parallel ($Q_x$) to the surface that are changed simultaneously as a function of the exit angle $\alpha_f$. 
Accordingly, the scattering signal reflected vertical density correlations in the z-direction ($Q_z$) of the ML structure, roughness correlations along the surface plane ($Q_x$) direction and cross-correlations between both~\cite{Holy93}. For reasons of clarity, we plot here only the $Q_z$ values but we note that the later in-depth density reconstruction analysis kept track of both $Q_x$ and $Q_z$ directions and interface correlations. The scattering signal at $Q_z > Q_{\text{specular}}$ is closely associated with the specular reflectivity curve~\cite{Holy93} which we characterised separately \textit{ex situ} (see \textit{Methods} and \textit{Supplementary Information}). The intense peak at $Q_z = 1.33 \,\mathrm{nm}^{-1}$ represents the typical length scale, \textit{i.e.} 15.8~nm thickness of each Ta/$\mathrm{Cu_3 N}$ double-layer. The Kiessig fringes \cite{Kiessig31}, represented as small peaks between $Q_z = 1.0 \,\mathrm{nm}^{-1}$ and $1.33 \,\mathrm{nm}^{-1}$, are a fingerprint of the number of double-layer repeats in the ML sample. Upon interactions with the laser, we observed drastic changes in the x-ray scattering profiles on a picosecond time scale. We notice a progressive reduction of the number of Kiessig fringes with time accompanied by a broadening and reduced intensity of the peak at $1.33 \,\mathrm{nm}^{-1}$ (Fig.~1~c) both indicating a receding surface and a loss of correlation within the double-layer structure. At time scales of $3-4 \,\mathrm{ps}$, the Kiessig fringes are almost completely vanished and only a broad peak of the basic double-layer structure remained visible. 
Especially instructive and important for the later density reconstruction is the scattering signal at exit angles below the incident angle ($Q_z < Q_{\text{specular}}$). Here the different exit angles are in the range of the critical angles of total external reflection of the respective materials where a surface sensitive evanescent x-ray wave travels parallel to the surface. These so-called Yoneda peaks \cite{Yoneda63} are thus originating predominantly from the interference of the top-most surface layers and are a sensitive marker for the structure of the uppermost layers. Accordingly, the peaks at $0.80 \,\mathrm{nm}^{-1}$ and $0.87 \,\mathrm{nm}^{-1}$ (dashed vertical lines Fig.~1~a) correspond to the solid densities of $\mathrm{Cu_3 N}$ and Ta, respectively. It is striking that, even at the delay time of $4 \,\mathrm{ps}$ where the Kiessig fringes had disappeared, we observed well defined Yoneda peaks, which facilitate a reconstruction of the electron density. It is worth mentioning that the analysis of the diffuse scattering signal and the Yoneda peaks is also referred to as grazing-incidence x-ray diffuse scattering~\cite{Sinha88}. We call the method GISAXS here as this is the most common term being used at synchrotrons~\cite{Pietsch04}. GISAXS has the advantage that depth information is provided without scanning incident and exit angles as required by x-ray reflectometry \cite{Holy93, Hexamer15}, so that it can be well adapted for single-shot XFEL experiments.

The information around the dynamical diffraction effects as well as the \textit{in}- ($Q_y = 0$) and \textit{out}- ($Q_y \neq 0$) of-plane data is used to reconstruct the density profile by making use of the state-of-the-art GISAXS analysis program BornAgain~\cite{Burle18}. This analysis is similar to Ref.~\cite{Schlomka95} and yields the temporal evolution of both the depth-resolved real-space density profile and the parameters describing the correlated roughness between the layers. A simultaneous refinement of in-plane and out-of-plane scattering signals was important to mitigate problems of non-unique solutions and key to a successful density reconstruction. s- (details in \textit{Supplementary Information}). Circular dots in Figs.~\ref{fig:fig2}~a-b) show the in-plane scattering profile obtained at $-0.5$ and 2.0~ps after the laser intensity peak, respectively, while Fig.~\ref{fig:fig2}~c) shows the corresponding retrieved electron density profile. The laser irradiates the sample from the left side. The higher and lower effective electron density layers represent Ta ($n_\mathrm{eff} = 3.7 \, \mathrm{nm^{-3}}$) and $\mathrm{Cu_3 N}$ ($n_\mathrm{eff} = 2.2 \, \mathrm{nm^{-3}}$), respectively. The shaded areas indicate a confidence interval of the electron density profiles which are determined as described in the \textit{Supplementary Information}. 
The solid lines in Fig.~\ref{fig:fig2}~a-b) represent fits through iteratively minimizing chi-square values resulting in the density profiles shown in Fig.~\ref{fig:fig2}~c). The $-0.5$~ps signal (\textit{i.e.}, the integration over $-0.5 \, \pm$ 0.6~ps around the laser intensity peak) is identical to the density profile in the cold sample. At 2.0~ps delay, the density profile is strongly modulated, although there still existed a periodic double-layer structure, as manifested by the continued presence of the peak at $Q_z = 1.33 \,\mathrm{nm}^{-1}$. A reduced number of layers or reduced homogeneity in layer thickness corresponds to the disappearance of Kiessig fringes between $Q_z = 1.0$ and $1.33 \,\mathrm{nm}^{-1}$. Further, the reduced density of the uppermost Ta layer in Fig.~2~c) is corroborated by the reduction of the $Q_z = 0.87 \,\mathrm{nm}^{-1}$ peak. Finally, the front of the expanding plasma is inferred from a new shoulder appearing at $Q_z = 0.73 \,\mathrm{nm}^{-1}$. 
Note that we track here mostly the bound electrons, or equivalently, the ion motion. This is because we probe here all electrons with binding energies below the photon energy, i.e. the \textit{effective} number of electrons per atom (27 for Cu and 63 for Ta), while the mean ionization predicted under our laser condition is relatively small (up to $Z_{mean} = 7~$\cite{Chung05}).
\begin{figure*}[htb!]
    \centering
    \includegraphics[width=\linewidth]{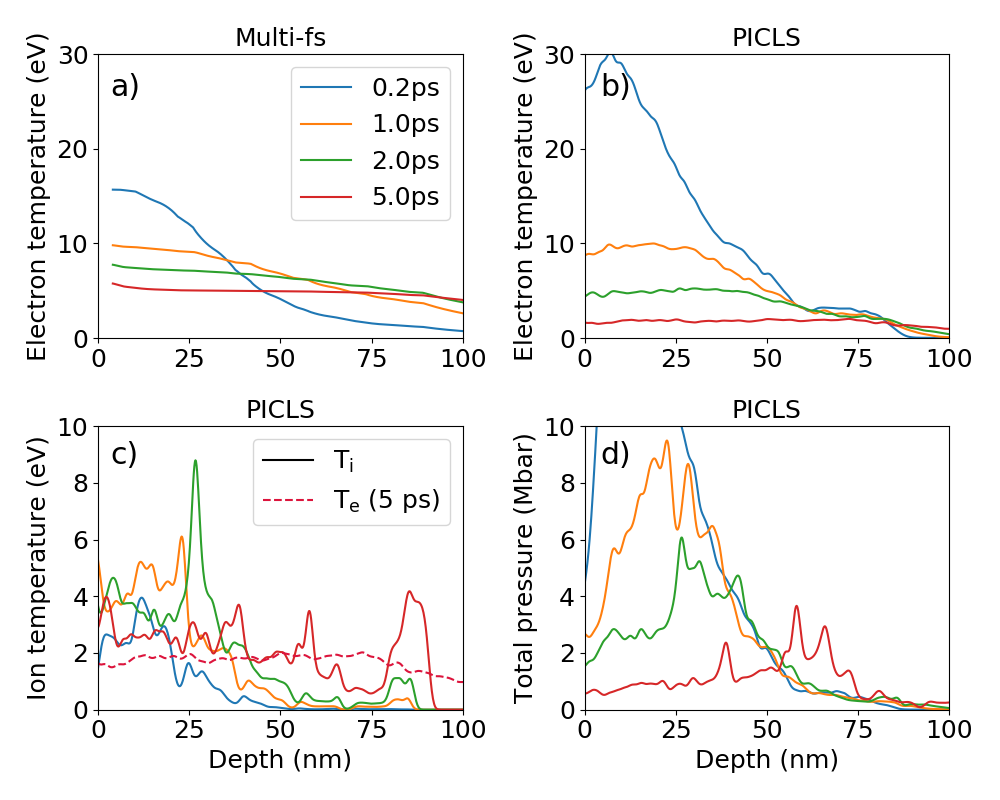}
    \caption{\textbf{Electron and ion temperatures, pressures inside the multilayer at 0.2, 1.0, 2.0, 5.0~ps delay.} Electron temperature $T_e$ vs. depth calculated by \textbf{a)} Multi-fs 1D hydrodynamics simulation and \textbf{b)} PICLS 1D kinetic simulation. \textbf{c)} Ion temperature $T_i$ vs. depth calculated by PICLS. The red dashed line is $T_e$ at 5~ps delay. \textbf{d)} Pressure (electron + ion) profile calculated by PICLS.  
    For all simulations, the laser intensity and the pulse duration are $4\times 10^{14} \mathrm{W/cm^{-2}}$ and $50$~fs (FWHM), respectively. The laser irradiates the sample from the left side. The depth of $z = 0$ corresponds to the initial solid surface. The laser intensity peak is at 0~fs.} 
    \label{fig:fig4}
\end{figure*}
\subsection{Comparison with simulations}
In a next step, we compare the experimentally retrieved density profiles with simulations for the different delays. In our experiment with $3.6 \times 10^{14} \, \mathrm{W.cm^{-2}}$ intensity, the laser absorption is dominated by inverse bremsstrahlung. Electrons are thermalized in femtoseconds via collisions and heat transport into the bulk is influenced by collisional ionization and degeneracy. As the dynamics of our system is located between the kinetic and hydrodynamic regimes, we used here two state-of-the-art simulation codes to represent these regimes by implementing the same collision model for strongly-coupled plasmas: 
\begin{enumerate}
    \item Multi-fs, a one-dimensional (1D) Lagrangian two-temperature hydrodynamics code, which is specifically designed for femtosecond laser-solid interactions in a non-relativistic regime \cite{Eidmann00, *Ramis12}. Multi-fs represents Lagrangian codes such as HYDRA and HELIOS~\cite{marinak2001three, *MACFARLANE06}, which are widely used for modelling nanosecond laser plasma interactions in the context of inertial confinement fusion (ICF)~\cite{Atzeni04}.
    \item PICLS, a kinetic collisional electromagnetic particle-in-cell (PIC) code \cite{SENTOKU2008} representing widely used codes such as EPOCH, VLPL and OSIRIS~\cite{arber2015contemporary, *pukhov1999three, *fonseca2002osiris}. PIC codes are typically used for the modelling of relativistic plasmas in the context of laser particle acceleration.
\end{enumerate}
For modelling warm and dense plasmas, the electron collision frequency $\nu$, which is dominated by the sum of electron-electron $\nu_{ee}$ and electron-ion $\nu_{ei}$ terms, must be properly implemented as collisions have a large impact on laser absorption, electron impact ionization, electron-to-ion energy transfer, and heat conduction -- processes that determine the plasma density dynamics. Since our plasma is located between the classical hot plasma regime ($T_e \gg T_F$) and the degenerate electron regime ($T_e \leq T_F$), where $T_F = 7.1$ and 8.6~eV is the Fermi temperature for Cu and Ta, respectively, we implemented an interpolation between two regimes, similar to what is implemented in Multi-fs \cite{Eidmann00, *Ramis12} (\textit{Methods}). The typical effective collision time ($\nu^{-1}$) in our regime appears about 0.1~fs with an electron mean free path length of below 1~nm. 
In Multi-fs, the microscopic velocity distribution at a given point in space is replaced by a local, averaged particle velocity and temperature. The missing microscopic individual particle dynamics in Multi-fs are incorporated in PICLS by introducing Monte-Carlo binary collisions and an extremely fine cell size of 0.125~nm. However, the implemented collision model only takes into account the small-angle binary collisions, while large-angle deflection and many body collisions are important in the regime of strongly-coupled plasmas. Our main goal here is to assess the predictive capability of these widely-used plasma simulation codes.\\ 

Fig.~\ref{fig:fig4} summarizes the temperature profiles obtained by a) Multi-fs and b) PICLS. 
It can be seen that the electron temperature at the surface region is instantaneously elevated to 15--30 eV with a spatial gradient inside the bulk of size $\sim$40~nm which agrees with the collisional skin depth $c/ \omega_{pe} \sqrt{\nu/2\omega_L}$, where $c$ is the speed of light, $\omega_{pe} = \sqrt{e^2 n_e / m_e \varepsilon_0^2}$ is the plasma angular frequency, $\omega_L$ is the laser angular frequency, and $e, \, n_e, \, m_e$ and $\varepsilon_0$ are electron charge, free electron density, electron mass and vacuum permittivity, respectively. The temperature gradient quickly dissipates by the thermal diffusion by conduction which can be described by the two-temperature energy conservation equations (\textit{Methods}). As a result, a uniform electron temperature profile is obtained at $\sim$2~ps as supported by both simulations. Here, heating of deeper layers via ballistic electrons plays a negligible role due to high collisionality, as corroborated by PICLS. 
Ions are slowly equilibrated with electrons via collisions from the front region, and the ML system is equilibrated within $\sim$5~ps (Fig.\ref{fig:fig4} c). The simulation results by PICLS show a higher initial temperature ($\sim$30~eV) compared to that of Multi-fs ($\sim$16~eV), but the PICLS showed a quicker temperature decay (discussion below). The mean ionization $Z_{mean}$ for the top Ta and Cu layers is about $5^+$ to $7^+$, respectively, in both simulations. \\

\begin{figure*}[htb!]
    \centering
    \includegraphics[width=\linewidth]{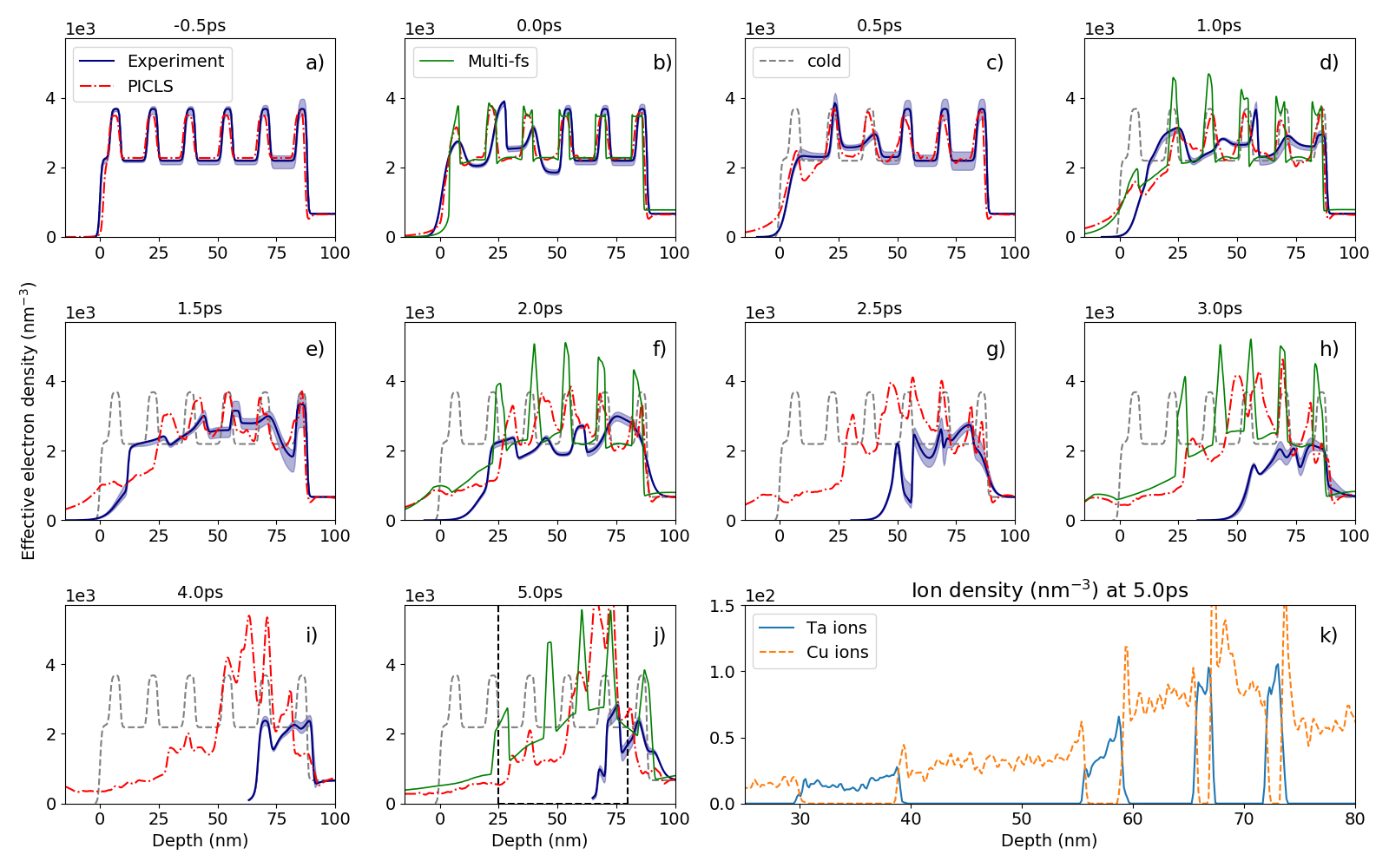}
    \caption{\textbf{Evolution of the effective electron density profile $n_{\mathrm{eff}}$ vs. depth with various pump-probe delays. Comparison between the experiment, kinetic and hydrodynamic simulations.} \textbf{a-j)} Time delay of -0.5 to 5~ps. Blue-solid lines are experimental results, red dash-dotted lines are simulation results from the 1D PICLS, and green-solid lines are results from the 1D Multi-fs. For better visibility, Multi-fs data is shown only for selected delays. Both simulation data are averaged over 1.2~ps to take into account the experimental time resolution. \textbf{k)} Ta (blue-solid) and Cu (orange-dashed) instantaneous (not time-averaged) ion density profile calculated with PICLS. The 25--80~nm area is zoomed for better visibility of layer intermixing, as indicated by the dashed box in j). 
    }
    \label{fig:fig3}
\end{figure*}
The electron density profiles from the experimental measurement, the Multi-fs, and the PICLS simulation are summarized in Fig. \ref{fig:fig3} in time steps of 0.5 ps. It can be seen that due to the strong localized surface heating, both experiment and simulations show immediate expansion of the top Ta layer into vacuum due to high surface pressures in excess of $>$ 10~Mbar (Fig. \ref{fig:fig4} d). The pressure peak propagates into the bulk, and its amplitude gradually decreases together with the temperature. Due to the fast thermal conduction, the whole ML becomes modulated after $\sim$1~ps. 
The ablation of the solid-density surface developed over time, with a speed of $\sim$10~nm/ps up to 2~ps in a good agreement with the experiment and with the calculated speed of sound (\textit{Methods}). 
The simulation results from Multi-fs (green-solid line in Fig. \ref{fig:fig3}) however show a density dynamics at the later times predicting that the structure of all layers except for the first one are essentially preserved for the entire time window under investigation, in stark contrast with the experiment.
We attribute this to the prohibited particle interdiffusion between Lagrangian cells used in the Multi-fs code. With the ion thermal velocity ($\sqrt{k_B T_i/M_i} \geq$ 1~nm/ps with $\sim$1~eV thermal temperature, where $k_B$ is the Boltzmann constant, $T_i$ is the ion temperature and $M_i$ is the ion mass), particles are expected to penetrate into adjacent layers over timescales of picoseconds. 
Such kinetic effects are included in PICLS which lead to a better agreement with the experimental results showing the significant modulation of all layers starting around 1~ps. 
Furthermore, the global shape of the density profiles and the number of density peaks observed in PICLS are in qualitative agreement with the experiment for the entire time window. 
However, after 2.5~ps, the amount of the volume left behind was overestimated in PICLS, indicating that the thermal energy contained in the bulk was underestimated.
The laser absorption into electrons is 45$\%$ in PICLS in agreement with Ref.~\cite{Price1995} indicating that collisional absorption was simulated reasonably well. On the other hand, it has been revealed that one-third of the total kinetic energy of particles is lost by the continuous radiation emission with a peak at $\sim$ 90~nm wavelength. 
The bremsstrahlung radiation in this wavelength range is intrinsically included in our PIC simulation via Maxwell's equations. However, the radiation transport and opacity was likely underestimated as the photoabsorption was excluded. 

Furthermore, three-body recombination is not included in the code, which overestimates the number of free electrons and therefore the bremsstrahlung emission at later times. 
In addition, as the temperature regime in our experiment is around the Fermi temperature, the electron distribution function should obey Fermi-Dirac statistics rather than Maxwell-Boltzmann statistics which is used in the simulation. This should affect both the collisional ionization rate and the electron energy distribution function \cite{Williams20}.
All of the above considerations indicate that the particle kinetic energies were most likely underestimated at later times, which support the slower density dynamics and the ablation speed observed in PICLS compared to the experiment. To correctly implement those dynamics into the code requires atomic and quantum physics models of strongly-coupled, partially ionized dense plasmas interacting with macroscopic ($\gg$nm) electromagnetic fields. This lies beyond the capability of currently available simulation codes and computing resources.

Another important aspect that has to be considered is the interpenetration of ion particles into adjacent layers. 
If the intermixing is fully excluded, as in the case of 1D Langrangian Mulfi-fs code, the boundary between two neighbouring layers acts as a piston pushing against each other driven by the instantaneous pressure of each layer. This produces oscillations of the boundary until the pressure equilibrium is reached~\cite{Gislason10}. Inner layers are tampered by the first layer until it is fully ablated, slowing down the whole ablation process. In parallel, the forward propagating pressure wave compresses inner layers, producing local density peaks as seen in both simulations. 
On the contrary, if the ion interpenetration into adjacent layers occurs, it smooths out the layer boundary, and the ablation would occur as a whole. 
In PICLS, although the particle interpenetration is included, we observe that the length scale of ion penetration was limited only to $\sim$1~nm even at 5~ps delay (Fig. \ref{fig:fig3} k), so that the layers are still well-separated. Our hypothesis is that the slow intermixing is caused by the 1D geometry in the simulation, resulting in a lack of multi-dimensional instabilities which would enhance the intermixing of the layers~\cite{Kluge15}. Unfortunately, a multi-dimensional kinetic simulation for a multi-ps time scale with {\AA}ngstr\"{o}m resolution cannot be performed due to its huge computational costs. 
To perform a crude estimate of the degree of intermixing, we performed a simplified 2D PICLS simulation with a reduced geometry (Cu/Ta/Cu 3 layers only) and without laser interaction but implementing a homogeneous initial electron temperature instead. The simulation revealed that the ion interpenetration was at least a factor of two faster in the 2D simulation (\textit{Supplementary Material}).

We further observed in the PICLS simulations that the collision frequency $\nu$ and the cell size strongly affected the laser absorption and subsequent density dynamics. When we reduced the electron-ion collision frequency $\nu_{ei}$ to 20$\%$ of that of the original value (while keeping the electron-electron collisions $\nu_{ee}$), the laser absorption was reduced to 16$\%$ from 45$\%$, due to the enhanced harmonic electron oscillation under the laser field which enhanced the reflection. A similar effect was observed when increasing the cell size from 1.25~$\mathrm{\AA}$ to 6~$\mathrm{\AA}$. The reduced spatial resolution underestimates the atomic-scale collisions, leading to decreased absorption to 34$\%$. These reductions of absorption slow down the density dynamics and cause the agreement with experimental results to be lost. Our analysis thus shows the importance of implementing the correct collision frequency and the ability to resolve atomic collisions in dense plasmas.

\section{Discussion}
This first visualization of ultrafast, subsurface solid-density dynamics of nano-plasmas upon intense laser irradiation has revealed several important findings. Surprisingly, widely-used hydrodynamic simulations lose their predictive power when particle interdiffusion between Lagrangian cells occurs. Although kinetic PIC simulations have been considered inadequate for low temperature, high-density plasmas in the past, here we observe that they show much better agreement with experimental results by implementing atomic-scale collisions. 
In particular, our study validates the code up to a few picoseconds after the laser irradiation: the surface ablation speed shows an excellent agreement with experiments and with the speed of sound, i.e. the speed of density depletion propagating into bulk driven by the expansion. The agreement is lost at later times due to the current limitations of the code: missing physics of radiation transport, multi-dimensional atomic mixing, recombination, large-angle many-body collisions and degeneracy. This will motivate the development of new models which can now be tested by using our experimental method. 
A particularly interesting direction in which to proceed is to use a hybrid approach, for example using an electromagnetic PIC to calculate the long-range electromagnetic force while using a classical electrostatic molecular dynamics to resolve the microscopic density fluctuations and atomic-scale collisions \cite{Varin2014, *Bart2017}.  
Although quantum effects such as Pauli blocking cannot be implemented in such simulations, a comparison with the experiment will allow an evaluation of the importance of these effects and validate models. Experimental GISAXS measurements of the surface and subsurface dynamics of dense plasmas will in turn allow the benchmarking of physics models and simulations with relevance to laser material processing and high-energy-density science. 

The presented GISAXS experiment can be improved by utilizing a larger detection area, which would enlarge the accessible $Q$-range and would thus provide additional constraints on the density retrieval. We have verified that the GISAXS signal is intense enough up to at least $Q = 2.5 \,\mathrm{nm}^{-1}$ for our experimental setup. Furthermore, the temporal smearing due to the grazing incidence, which limited our temporal resolution to be 1.2~ps, can be mitigated by using a smaller x-ray focal spot, \textit{e.g}, 100~nm, which is readily available at XFEL facilities \cite{Schropp15}. This would lead to only $\sim$30~fs smearing at 0.64$\degree$ grazing incidence, allowing experimental studies on surface and subsurface dynamics with nanometer depth and femtosecond temporal resolution. 

In addition to the depth sensitivity along the $Q_z$ direction, GISAXS also contains information on the surface and interface roughness, and their correlations between layers by analyzing along the $Q_y$ direction (see \textit{Supplementary Information}) \cite{Holy93}. With this, insights into the nanoscale ablation dynamics (via $Q_z$) and its role in the change of surface and interface roughness ($Q_y$) can be obtained simultaneously. Ultrafast GISAXS may thus resolve the generation of nano-gratings, so-called laser-induced periodic surface structures (LIPSS)~\cite{Bonse20}, which have applications for antibacterial, optical, chemical and industrial purposes ~\cite{Lutey2018, *San-Blas_LIPSS_20, *Li_LIPSS_20, *Bonse_Tribology18}. 

Furthermore, short-wavelength perturbations of the ablation surface has been found to seed atomic scale material mixing and has been investigated as a primary factor for reduced thermonuclear reaction rates in ICF experiments \cite{Ma13, *Haines2020}. A correlation of surface roughness and material mixing, growth of ripples \cite{Sgattoni15} and surface instabilities \cite{Kluge15} and their impact on ablation is thus a ubiquitous problem in light matter interactions. The high resolution and sensitivity of ultrafast GISAXS is expected to make a decisive contribution for a better understanding and control of these phenomena.]

\section{Methods}
\subsection{Experimental setup, x-ray and laser parameters and timing synchronization}
\noindent The experiment was performed at the SACLA XFEL facility in Japan at the Experimental Hutch 6 (EH6) which features a high-intensity optical laser ($\lambda_L=800 \,\mathrm{nm}$ central wavelength, maximum 10~J/pulse with 40~fs duration in full-width half-maximum (FWHM)) combined with ultrashort intense x-ray pulses \cite{Yabuuchi19}. The x-ray pulses had a photon energy of 8.81~keV (with 43~eV FWHM bandwidth), $\sim 100 \,\mathrm{\mu J/pulse}$, and a pulse duration of 7~fs in FWHM. The x-rays were focused to a $4 \,\mathrm{\mu m}$ FWHM spot on sample by a set of compound refractive lenses placed 3~m upstream from the sample. The scattered x-ray signal was recorded on a MPCCD area detector with $50 \,\mathrm{\mu m}$ pixel size \cite{Kameshima14} placed at a distance of 1.27~m from the sample and shielded by a $50 \,\mathrm{\mu m}$ thick Al foil, to remove plasma-induced bremsstrahlung background. The incident angle was fixed at $0.64 \degree$ to be slightly larger than the critical angle of total external reflection $\alpha_c \varpropto \sqrt{n_e}$ for both Ta  ($\alpha_{\text{c\_Ta}} = 0.46 \degree$) and Cu ($\alpha_{\text{c\_Cu}} = 0.35 \degree$). The x-ray photon energy was chosen not to overlap with any absorption edges or resonant lines of both Cu and Ta. The closest edges were 8.98~keV for the Cu K-edge and 9.88~keV for the Ta L-III edge. When the ionization develops with temperature, bound-bound absorption may appear: namely the 1s-3p transition in Cu around 8.95~keV at $T_e \geq 75$~eV and 2s-4f transition in Ta around 9.7~keV at $T_e \geq 90$~eV \cite{Chung05}. Nevertheless, these transitions are still far enough from the x-ray photon energy. Therefore, we can safely neglect the dispersion correction terms for our analysis. The x-ray probe was sensitive to the electrons with binding energies below the photon energy, i.e., 27 and 63 electrons for Cu(29) and Ta(73), respectively (total electrons in parenthesis). The sample was irradiated by a high-intensity optical laser attenuated to deliver $\sim 53 \,\mathrm{mJ}$ energy/pulse, impinging on the sample at $17 \degree$ incident angle from the surface normal with p-polarization. In order to cover the x-ray footprint on sample ($4 \,\mathrm{\mu m}$ FWHM for $0.64 \degree$ grazing incidence yields $360 \,\mathrm{\mu m}$), the optical laser beam was defocussed to a diameter of $\sim 500 \,\mathrm{\mu m}$ yielding an average laser intensity of about $3.6 \times 10^{14} \,\mathrm{W⁄cm^2}$ (see \textit{Supplementary Information}). The pump-probe delay timing was determined by the GISAXS pattern. As our x-ray scattering was an integration over $\sim$1.2~ps, the surface density should be modulated already at $t = 0$ (-0.6 -- +0.6~ps), as confirmed by PICLS. We set the delay to -0.5~ps for the data which showed the intact density profile. Independently, the temporal synchronization was measured a few hours before the experiment. Our delay appeared 1~ps earlier than this measurement, which we attributed mainly to the long-term timing drift \cite{Yabuuchi19}, and partially to the precision of the delay determination and the sample positioning accuracy. 

\subsection{Multilayer sample and its characterization }
\noindent The multilayer (ML) sample was prepared by DC magnetron sputtering at the University of Mainz. Five repeated layers of Ta and $\mathrm{Cu_3 N}$ were grown onto a thick silicon wafer carrying a Ta seed layer on a 100~nm thick layer of thermal silicon oxide, yielding a ML structure of, from the laser irradiation side, 5 repeat of Ta(4.3~nm thick)/$\mathrm{Cu_3 N}$(11.5~nm), Ta(4.3 nm), $\mathrm{SiO_2}$(100~nm) and Si substrate($700 \,\mathrm{\mu m}$). The wafer was then laser-cut into $4 \times 7 \,\mathrm{mm^2}$ individual pieces. Each sample was mounted on a rotation wheel which was individually pre-aligned with an attenuated x-ray FEL beam. Then we fixed the grazing incident angle to $\alpha_i = 0.64 \degree$ corresponding to the intense first ML Bragg peak at $Q_z = 1.0 \,\mathrm{nm^{-1}}$. This geometry, together with the size of the vacuum window and of the detector, allowed us to cover a $Q$-range of up to $1.5 \,\mathrm{nm^{-1}}$. The strong specular signal at the exit angle $\alpha_f = 0.64 \degree$ is blocked by a 3 mm diameter tungsten beam stop. The high visibility of the GISAXS signal indicates that our ML sample has a high degree of vertical correlation between the interfaces. The samples were pre-characterized at Technical University Dortmund with x-ray reflectometry using a $\theta - 2\theta$ reflectometer \cite{Krywka07}. For the measurement, the sample was mounted on an angle-dispersive reflectometer for angle adjustment between the sample and detector circle. This allows us to characterize the exact layer thickness (from the angular difference between the Kiessig oscillations), number of repeated layers (from the number of Kiessig fringes), surface roughness (from the overall reflectivity amplitude), and the average density of a layered system (from the critical angle of total external reflection, $\alpha_c$). Additional description as well as measured reflectivity curve is in the \textit{Supplementary Information}, Fig.~S4. The retrieved surface and interface roughness was $3 - 6 \,\mathrm{\AA}$ (root-mean-square value) depending on each layer.

\subsection{Hydrodynamic and PIC simulations}
\noindent 
We used a one-dimensional hydrodynamic simulation Multi-fs \cite{Eidmann00, *Ramis12} specifically designed for short ($\leq$ ps) pulse high-intensity laser-solid interactions below $< 10^{17} \,\mathrm{W/cm^2}$ laser intensity. 
The code calculates an explicit solution for Maxwell’s equations for the interaction of a laser pulse with a steep plasma density gradient, and includes a temperature-dependent collision frequency as well as a thermal conductivity from metallic solid up to high-temperature plasma, with separate equations for electron and ions (two-temperature model). 
In order to generate an equation-of-state (EOS) table, we used the FEOS code \cite{Faik18} based on a QEOS description \cite{More88}. The ionization state is obtained by the SNOP atomic code \cite{Eidmann94} for Ta, and by FEOS for Cu within the Thomas-Fermi description. 1020 cells were used for simulating the target which includes 820 cells for the multilayers (82~nm thick) and 200 cells for the $5 \,\mathrm{\mu m}$ thick substrate, respectively. To reduce the numerical error, we used finer cells closer to the surface and interfaces and wider cells in the middle of each multilayer. For the substrate material, we use aluminium (Al) instead of silicon. Radiation transport module was switched off. \\
The electromagnetic Particle-in-Cell (PIC) simulation was performed using the collisional 1d3v (one-dimensional in space and three-dimensional in velocity) PICLS code \cite{SENTOKU2008}. In order to resolve the microscopic particles dynamics with atomic collisions, the cell size was set to be $\Delta x = 1.25$~\AA~corresponding to the time--step $\Delta t = 4.16 \times 10^{-19}$~s. Each cell contains 15 virtual ion particles with an initial charge--states of 2, 1, and 3 for Ta, Cu, and Al, respectively. We also used the fourth--order particle--shape and different particle--weightings for different ion species. The ion number densities are set to realistic densities of Ta ($n_{\mathrm{Ta}}=31.8n_c$), Cu ($n_{\mathrm{Cu}}=48.7n_c$), and Al ($n_{\mathrm{Al}}=34.6n_c$), where $n_c = m_e \omega_L^2/(4 \pi e^2) = 1.742 \times 10^{21} \mathrm{cm^{-3}}$ is the critical plasma density at the laser wavelength of $\lambda_L = 800$~nm. Here, $m_e$ and $e$ are the electron mass and charge respectively, and $\omega_L$ is the laser angular frequency. The ionization dynamics are modeled using the field and direct-impact ionization models. To deal with the collisions at the electron temperatures ($T_e$) around the Fermi temperature ($T_F$), similar to Multi-fs, a model with an interpolated collision frequency was implemented. The incident angle of the laser in 1D PICLS is normal to the surface. 

\subsection{Collision frequency, velocities, electron mean free path}
\noindent The energy of electrons in an oscillating electric field, or the ponderomotive potential is $e^2 E_L^2⁄(4 m_e \omega_L^2) = 9.3 \times 10^{-14} I_L \lambda_L^2$, with $I_L$ and $\lambda_L$ in $\mathrm{W/cm^2}$ and $\mathrm{\mu m}$, respectively. At $3.6 \times 10^{14}  \,\mathrm{W⁄cm^2}$, this energy is 21~eV. The collision frequency $\nu$ is one of the most important physical quantities to describe the energy absorption and transport. The precise value of $\nu$ around the Fermi temperature ($T_F = \hbar^2 / 2 m_e \cdot (3\pi^2 n_e)^{2/3}$) is as of yet unknown, and interpolated formulae between metal-like solids ($\nu_\text{e-phonons} \varpropto T_e$) and ideal gas plasmas ($\nu_{ei} \varpropto T_e^{-1.5}$) are used in Multi-fs \cite{Eidmann00, *Ramis12}, which we also implemented in PICLS. The maximum value of $\nu$ was set such that the electron mean free path does not go below the ion sphere radius $R_0 =(4\pi n_e/3Z)^{-1/3}$ to avoid a non-physical behavior. Around $T_F$, its value is being close to the plasma frequency $\omega_{pe} \sim 0.1 \, \mathrm{fs^{-1}}$. The implemented collision frequency into PICLS is shown in the \textit{Supplementary Information}, Fig. S11.  
The velocity of individual electrons around or below $T_F$ in a metal is $v_F = \hslash (3 \pi^2 n_{e})^{1⁄3}⁄m_e$, which is 1.7 and 1.6 $\mathrm{\mu m⁄ps}$ for $\mathrm{Ta^{2+}}$ and $\mathrm{Cu^{1+}}$, respectively ($\hslash$ is the Planck’s constant). At $T_e > T_F$, the thermal electron velocity $v_{th} = \sqrt{k_B T_e⁄m_e}$ needs to be considered, which leads to an electron velocity of $v_e = \sqrt{v_F^2 + 3 v_{th}^2}$. At $T_e = 20 \,\mathrm{eV}$, it leads to $v_e$ = 3.7 and 3.6 $\mathrm{\mu m⁄ps}$, respectively. A corresponding electron mean free path is $v_e/\nu \sim 0.1-0.2 \,\mathrm{nm}$, which is significantly smaller than the skin depth. Under these conditions, ballistic transport of electrons is effectively suppressed and the diffusion approximation for the electron heat transfer can be safely assumed. 

\subsection{Heat diffusion}
\noindent
Because the temperature decays quickly after the laser pulse, the compression wave cannot overtake the heat diffusion. The dominant energy transport is driven by the heat diffusion that is expressed by the two-temperature ($T_e$, $T_i$ for electrons and ions, respectively) energy conservation equations: 
\begin{equation}
 \begin{split}
 & C_e \frac{\partial T_e}{\partial t} = -\nabla \cdot \bm{q} - \gamma(T_e - T_i) + Q(z,t),\\
 & C_i \frac{\partial T_i}{\partial t} = \gamma(T_e - T_i),
 \end{split}
 \label{eq:heat_diffusion}
\end{equation}
where $C_e$, $C_i$ are, respectively, the volumetric heat capacity of electrons and ions, $\bm{q} = \kappa \nabla T_e$ describes the electron thermal heat flow with $\kappa(\nu)$ being the heat conductivity, and $Q(z,t) = \partial I_{abs}⁄\partial z$ is the power density deposited by the laser with $I_{abs}$ being the absorbed laser flux. The energy transfer rate from electrons to ions is expressed by $\gamma = C_i \tau_i^{-1}$ with $\tau_i = m_i⁄(2 m_e \nu_{ei})$ being the characteristic time for ion heating, where $m_i$, $m_e$ are the electron and ion masses, respectively. With the hydrodynamic simulation, the position of the heat wave front can be defined as $1/e$ of the surface temperature at each time frame. From that, we obtain the heat wave speed of $\geq$ 60 nm/ps (\textit{Supplementary Information} Fig.S10), which shows a good agreement with the fact that whole ML starts to be modulated within $\sim$1~ps after the laser pulse . 

\subsection{Sound velocity}
\noindent 
The sound velocity is the speed of an ion-density modulation driven by the pressure wave. Assuming the ideal gas equation of state (EOS), the sound velocity can be expressed as,
\begin{equation*}
    C_s = \sqrt{\frac{\gamma_e Z_{mean} k_B T_e + \gamma_i k_B T_i}{M_i}},
\end{equation*}
where $Z_{mean}$ is the mean ionization, $\gamma_e$ and $\gamma_i$ are adiabatic index of electrons and ions, respectively and $M_i=181$ a.u. and 63 a.u. is the ion mass for Ta and Cu, respectively. 
From the ideal gas EOS, $\gamma = 1+2⁄n$ with $n$ being the number of degrees of freedom. 
In most cases, $\gamma_e = 1$ and $\gamma_i = 3$ can be used. 
As an example, for ($T_e$,$T_i$)=(20,5)~ eV and $Z_{mean} = 5$, $C_{s \_ Ta} \sim 8 \, \mathrm{nm⁄ps}$ and $C_{s \_ Cu} \sim 13 \, \mathrm{nm⁄ps}$, respectively.

\subsection{Acknowledgements}
\noindent
The XFEL experiments were performed at the BL2 of SACLA with the approval of the Japan Synchrotron Radiation Research Institute (JASRI) (Proposal No. 2018B8049). 
C.G. acknowledges funding by DFG GU 535/6-1. 
G.J, M.V-K and M.K acknowledge support by the EU (FET s-Nebula \#863155) and the ERC (SyG 3D MAGiC \#856538).
A.K acknowledges support by JST-Mirai Program Grant Number JPMJMI17A1, Japan.
C.R. is supported by the LOEWE excellence initiative of the state of Hesse..
Y.S. is supported by the JSPS KAKENHI JP19KK0072.
This work was partially funded by the Center of Advanced Systems Understanding (CASUS) which is financed by Germany’s Federal Ministry of Education and Research (BMBF) and by the Saxon Ministry for Science, Culture and Tourism (SMWK).”
We acknowledge invaluable input from C. B\"{a}htz (HZDR) for sample alignment and data interpretation. 
M.Na. thanks R. Kodama (U. Osaka), Th. Tschentscher, U. Zastrau (EuXFEL) and M. Yabashi (RIKEN) for their useful advice. 
M.Ba and M.Na thank E. Brambrink (EuXFEL) for valuable advice for hydrodynamic simulation. 
We thank A. Pelka (HZDR), and J-P. Schwinkendorf (EuXFEL) for various discussions. 
We acknowledge the support of M. Spiwek (DESY) and DISCO HI-TEC EUROPE GmbH for cutting ML samples. 

\subsection{Author Contributions}
\noindent
C.G. and M.Na. conceived the study. L.R., M.Ba., T.R.P., M.M., N.P.D., S.G., T.M., M.Ni., C.G., and M.Na performed the experiment at SACLA with support from T.Y., Y.I., K.S., and T.T. L.R. analyzed the GISAXS data under the supervision of C.G. M.Ba performed Multi-fs and PICLS simulations to interpret the data with extensive support of L.H and T.K, under the supervision of A.P.M., M. Bu., T.E.C and M.Na. N.P.D. and J.K.K. gave support for generating EOS tables. Multilayer samples were prepared by G.J., M.V-K., and M.K. which were further characterized by L.R., F.S., D.K., and M.P. before the experiment. J.K., M.M. and M.Na organized the sample cutting. L.R, M.Ba, C.G, and M.Na wrote the initial draft which was extensively revised by contribution from T.R.P., N.P.D., L.H., C.R., T.K. and A.P.M. The paper was further improved by input from T.Y., M.M., G.J., J.K.K., M.Ni., M.K. and C.F-G. All authors commented on the manuscript.

\subsection{Data Availability}
\noindent
The datasets generated during and/or analysed during the current study are available from the corresponding author on reasonable request. Most of data generated or analysed during this study are included in its supplementary information files. 

\subsection{Code Availability}
\noindent
The codes used during the current study are available from the corresponding author on reasonable request. 

\subsection{Additional information}
\noindent
\textit{Supplementary Information} is available in the online version of the paper. Reprints and permissions information is available online at xxx. Correspondence and requests for materials should be addressed to C.G. and M.Na.
\bibliography{Ref_GIXDS}
\end{document}